\documentclass[multphys,vecphys]{svmult}
\usepackage{makeidx}
\usepackage{graphicx}       
\usepackage{multicol}        
\makeindex
\begin{document}
\title*{The Growth of Black Holes and Bulges at the Cores of Cooling Flows}
\author{D. A. Rafferty\inst{1}, B. R. McNamara\inst{2}, P. E. J. Nulsen\inst{3}, \and M. W. Wise\inst{4}}
\institute{Ohio University,
\texttt{rafferty@ohio.edu}
\and University of Waterloo,
\texttt{mcnamara@uwaterloo.edu}
\and Harvard-Smithsonian Center for Astrophysics
\and University of Amsterdam}
\maketitle
\section{Introduction}
The intracluster medium (ICM) at the center of a majority of galaxy clusters has a cooling time less than $10^{10}$ yr \cite{pere98}. In the absence of a source of heat, this gas should cool, resulting in a slow inward flow of material know as a ``cooling flow''. However, recent high-resolution X-ray spectra from XMM-\textit{Newton} do not show the features expected if large amounts of gas are cooling below $kT \sim 2$ keV \cite{pete03}. The emerging picture of cooling flows is one in which most of the cooling is roughly balanced by heating from the active galactic nucleus (AGN) due to accretion onto the central supermassive black hole (BH). In this regulated-cooling scenario, net cooling from the ICM would lead to condensation of gas onto the central galaxy, driving the star formation observed in many systems (e.g., \cite{mcna89}). We test this scenario using star formation rates (SFRs), ICM cooling rates, and AGN heating and BH growth rates for a sample of cooling flows with AGN-created X-ray cavities.
\section{BH Growth, Cooling, and Star Formation}
Cavities seen in the X-ray emission of clusters allow a direct measurement of the non-radiative energy output via jets from the AGN. The total energy required to create a cavity is equal to its enthalpy ($E_{\rm{cav}}=\gamma/[\gamma - 1] pV$), which depends on the pressure ($p$) and volume ($V$) of the surrounding gas and the ratio of specific heats ($\gamma$) of the gas filling the cavity. The jets are powered by accretion, through the conversion (with efficiency $\epsilon$) of the gravitational binding energy of the accreting material into outburst energy. Since some of the accreting material's mass goes to power the jets, the BH's mass grows by $\Delta M_{\rm{BH}} = (1-\epsilon)E_{\rm{cav}}/\epsilon c^2$ during the outburst.

\begin{figure}
\centering
\includegraphics[height=5cm]{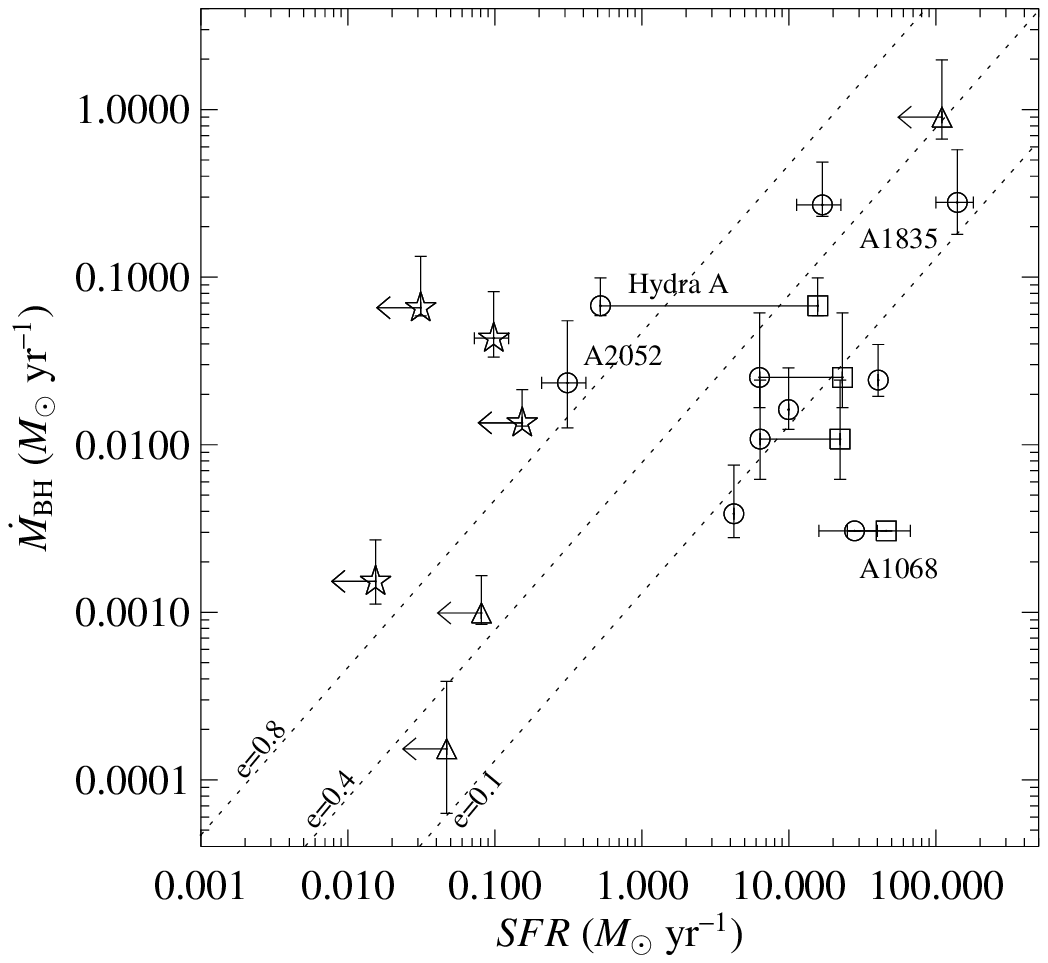}
\includegraphics[height=5cm]{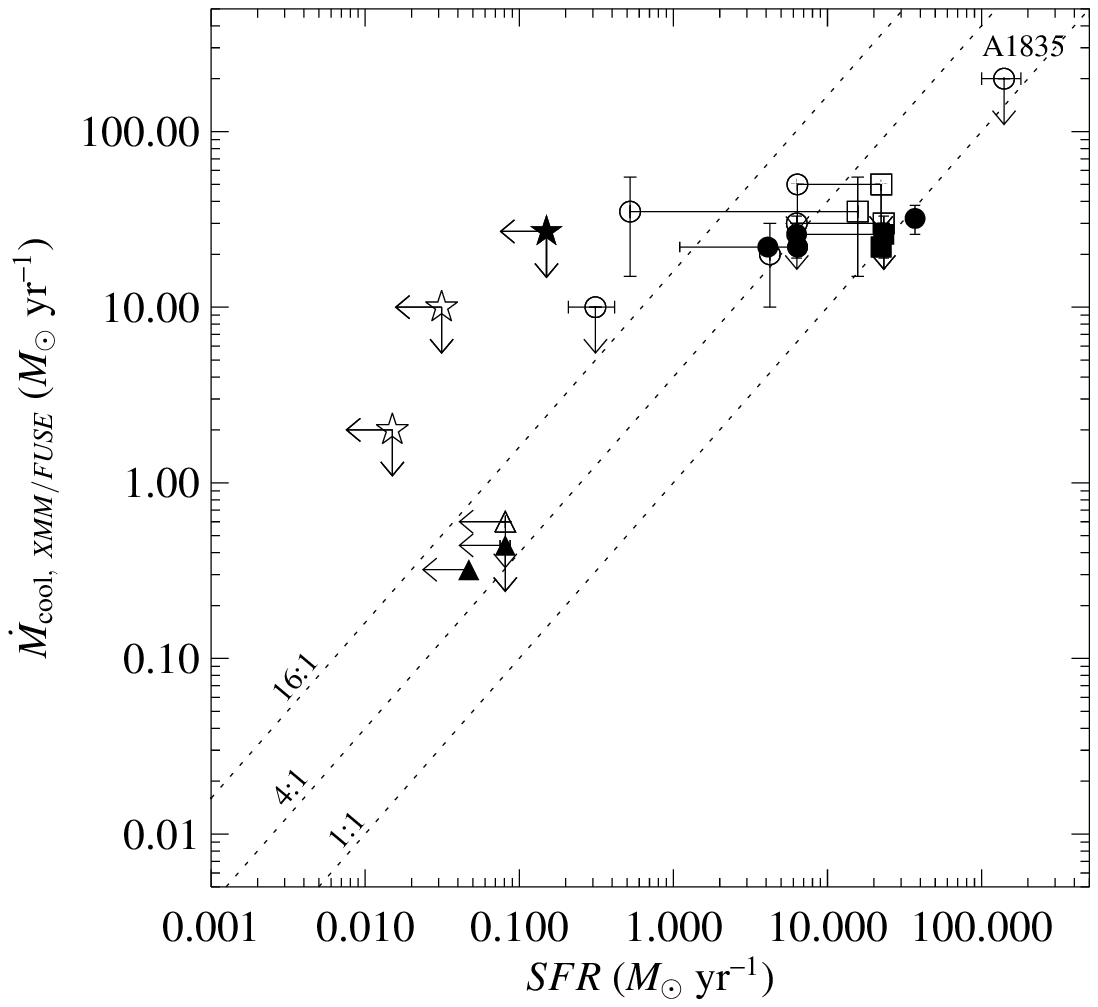}
\caption{\emph{Left:} Black hole versus bulge growth rate (SFR). Lines show the time derivative of the Magorrian relation (from \cite{hari04}) for various efficiencies ($\epsilon$).  \emph{Right:} Cooling rate versus SFR. Lines denote ratios of cooling to SFR, filled symbols denote \emph{FUSE} rates, and open symbols denote XMM rates. In both plots, circles and stars denote continuous SFRs derived from optical images or spectra, respectively; squares denote burst SFRs. Triangles denote far-infrared rates.}
\label{fig:growth_rates} 
\end{figure}
Figure \ref{fig:growth_rates} (\emph{left}) shows the BH growth rate (found by dividing $\Delta M_{\rm{BH}}$ by the cavity's buoyant rise time) versus the bulge growth rate (traced by star formation rates taken from the literature, see \cite{raff06} for details) for the systems in our sample with reliable star formation rate estimates. The trend in Figure \ref{fig:growth_rates} may indicate that, in a time-averaged sense, the growth of the bulges and BHs in our sample proceeds roughly along the Magorrian relation. The scatter indicates that present-day growth is occurring in spurts, with periods star formation (as in A1068\index{A1068}) in which the bulge grows quickly with little commensurate BH growth, while during periods of heating (as in Hydra A\index{Hydra A} or A2052\index{A2052}) the BH grows more quickly than the bulge. A1835\index{A1835} is a system in which the rates follow the Magorrian relation.

In the classical cooling flow problem, the X--ray-derived cooling rates were factors of $10-100$ in excess of the star formation rates in most systems.  Figure \ref{fig:growth_rates} (\emph{right}) shows that the star formation and cooling rates have converged greatly and are in rough agreement in some systems (e.g., A1835\index{A1835}). However, until line emission that is uniquely due to cooling below $\sim 2$ keV is identified, cooling through this temperature at any level cannot be confirmed.

\begin{figure}
\centering
\includegraphics[height=5cm]{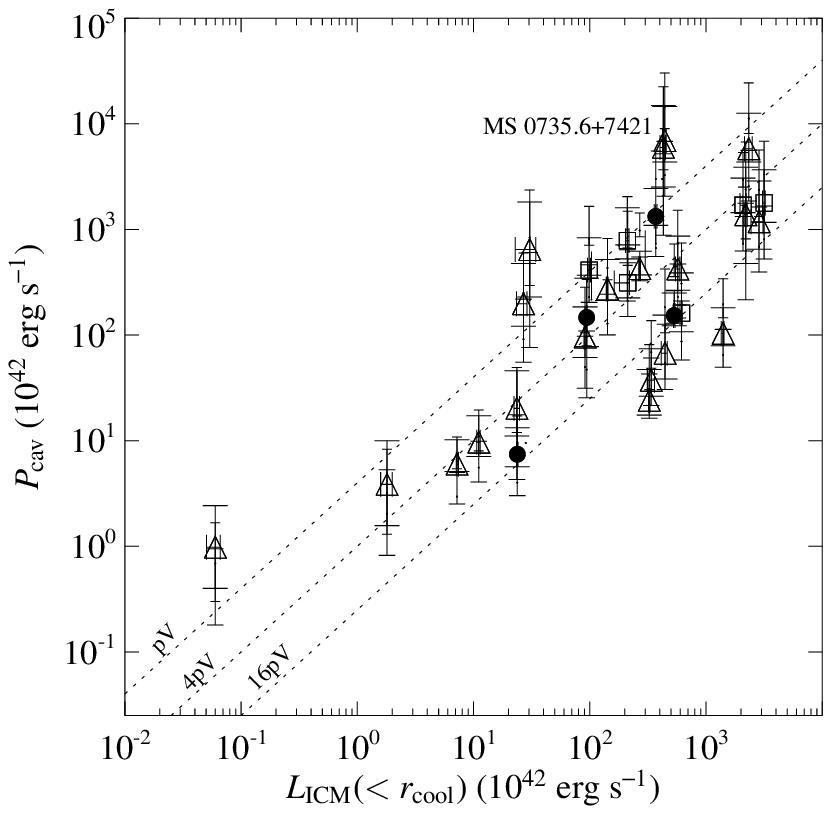}
\includegraphics[height=5cm]{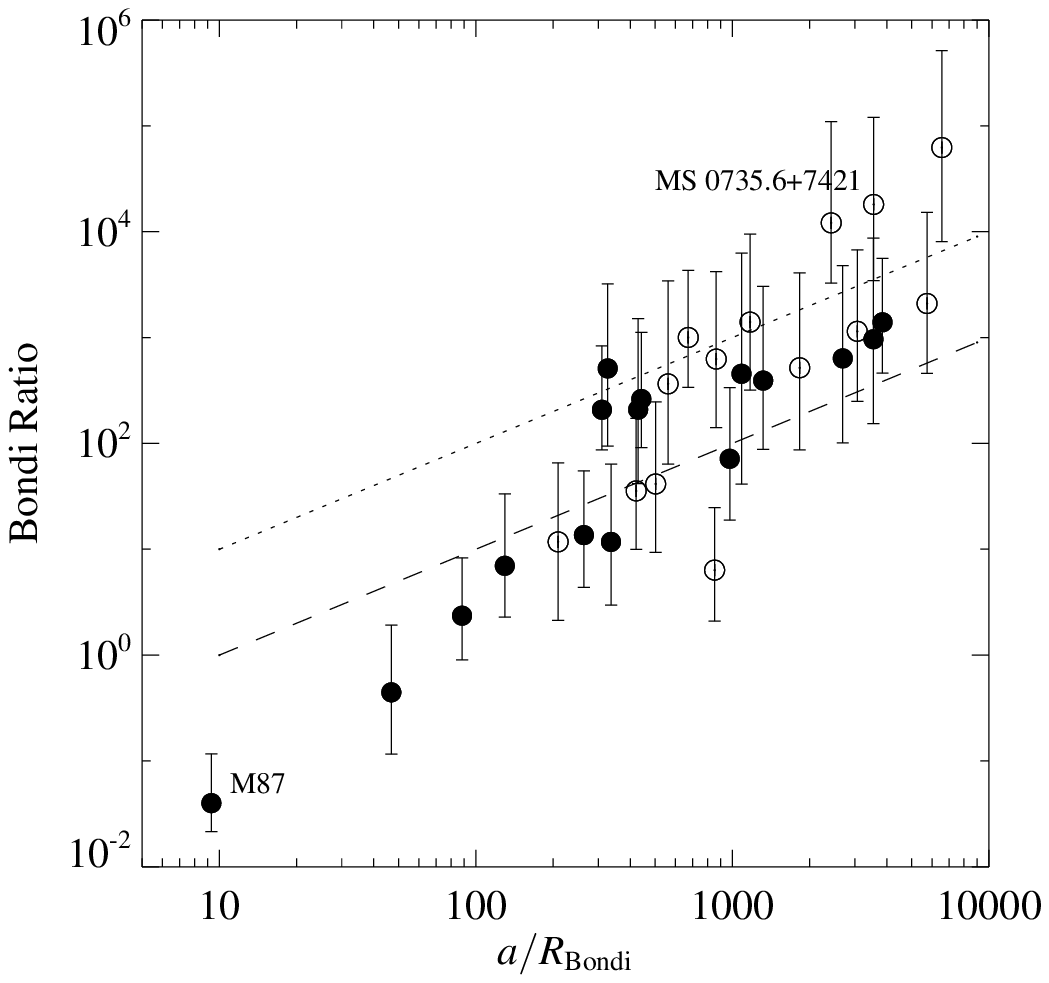}
\caption{\emph{Left:} Cavity power versus cooling luminosity. Lines denote equality for different values of $\gamma$. Circles, triangles, and squares denote well-defined, intermediate, and poorly-defined cavities, respectively. \emph{Right:} Bondi ratio versus size of the extracted region normalized to the Bondi radius. Lines denote the scaling of the measured Bondi ratio with radius, assuming a true Bondi ratio of unity at the Bondi radius and a density profile that rises as $\rho \propto r^{-1}$ to the Bondi radius (upper line) or flattens inside $a/R_{\rm{Bondi}}=10$ (lower line), as observed in M87\index{M87} \cite{dima03}.}
\label{fig:cav_power} 
\end{figure}

To investigate whether the AGN outbursts are powerful enough to balance cooling, we plot in Figure \ref{fig:cav_power} (\emph{left}) the cavity power of the central AGN against the luminosity of the ICM within the cooling radius. Remarkably, half of the systems in our sample have cavity powers sufficient to balance the entire radiative losses of the ICM within the cooling radius (as found ealier by \cite{birz04}).  However, we note that the time-dependent nature of AGN feedback does not require that cooling is always balanced by heating \cite{omma04}.

Finally, we plot in Figure \ref{fig:cav_power} (\emph{right}) the ratio of the accretion rate to the Bondi rate versus the semi-major axis of the central region from which the Bondi rates were calculated. Objects near or below the overplotted lines could reasonably have ratios of order unity or less and thus be consistent with Bondi accretion. While Bondi accretion can feed the less powerful outbursts easily (see also \cite{alle06}), the most powerful outbursts (such as in MS0735.6+7421\index{MS0735.6+7421}) are generally inconsistent with Bondi accretion. 

\section{Conclusions}
We have presented an analysis of the star formation and AGN properties in the central galaxies in the cores of cooling flows (discussed in detail in \cite{raff06}). We find that AGN outbursts in most systems with cavities have enough energy to offset much of the radiative losses of the ICM, and to severely reduce cooling to levels that approach the star formation rates in the central galaxy. Using the cavities to infer BH growth and star formation to infer bulge growth, we find that bulge and BH growth rates scale with each other on average in accordance with the slope of the Magorrian relation, but with large scatter that may indicate that growth occurs in spurts.

\end{document}